\documentstyle[aaspp4]{article}

\begin{document}

\def\simgt{\hbox{\rlap{\raise 0.425ex\hbox{$>$}}\lower 0.65ex\hbox{$\sim$}}}
\def\simlt{\hbox{\rlap{\raise 0.425ex\hbox{$<$}}\lower 0.65ex\hbox{$\sim$}}}

\title{Weak Lensing Induced Correlations between 1Jy QSOs and APM galaxies on  Angular Scales of a Degree }

       \author{Dara J. Norman}
       \affil{University of Washington \\ Department of Astronomy \\ Seattle, WA 98195-1580}
       \authoraddr{Department of Astronomy, Seattle, WA 98195-1580}
       \and 
       \author{Liliya L.R. Williams}
       \affil{University of Victoria\\ Department of Physics and Astronomy \\  Victoria, BC, V8P 1A1, Canada}
       \authoraddr{Canada}

\begin{abstract}
We find angular correlations between high redshift radio selected QSOs 
from the 1 Jy Catalog and APM galaxies on $\simlt 1^\circ$ scales.
We demonstrate that observed correlations are 
qualitatively consistant with a gravitational lensing explanation and
are inconsistant with a Galactic dust obscuration model. Comparing our
results with those of Ben\'{\i}tez \& Mart\'{\i}nez-Gonz\'{a}lez, 
who also use 1 Jy
sources and APM galaxies we come to the conclusion that galaxy selection
criteria can have a major effect on the angular scale and amplitude of
detected correlations.  
\end{abstract}

\section{INTRODUCTION}

Large-scale matter inhomogeneities at low redshifts ($z \simlt 0.5$)
can have a profound effect on how we perceive the high redshift
universe.  Weak gravitational lensing is expected to produce statistical 
association of background QSOs and foreground galaxies due to a 
phenomenon known as magnification bias (Turner 1980, Canizares 1981). 
Magnification bias arises because gravitational lensing changes the 
solid angle of a source but conserves its surface brightness. Faint 
background objects are brightened into a flux limited sample while their 
number density on the sky is diluted. These two effects lead to opposite 
results in the observed number density of the sources.  If the slope of 
the magnitude-number counts is steep enough such that the enhancement of the 
number density due to source brightening `wins' over the geometrical 
dilution then an overdensity is observed. The opposite is true if the
source number counts are shallow. Any single object is not expected to 
confirm the presence of magnification bias because the large-scale
lensing effect is weak, the intrinsic luminosities of individual sources
are not known, and the number density variation of galaxies due to 
large-scale structure is large. However, a large enough sample of sources 
with  steep number counts will show an effect of magnification
bias in a statistical sense: an overdensity of sources behind lenses
will be observed.

Several studies have detected statistical associations of bright
background QSOs with foreground galaxies, believed to be tracers of
the lensing mass (e.g. Tyson 1986, Bartelmann \& Schneider 1997,
Norman \& Impey 1999, and references within). The scale over which an 
enhancement of galaxies is seen is important to the determination of the 
size of the foreground lens structures, as the angular scale of 
association can be translated to the physical scale when the redshifts 
of the lenses (galaxies, clusters) are known. Most studies have focused 
on scales $< 15'$, or $< 2h^{-1}$Mpc at the redshifts of typical lenses.
In these cases, when the slope of the number counts 
of the source sample has been steep enough $(d$log$N/dm=\alpha > 0.4$),
positive correlations have been detected (Fugmann 1990, 
Ben\'{\i}tez \& Mart\'{\i}nez-Gonz\'{a}lez 1995, 1997).  
In the case where the slope has been $< 0.4$, anti-correlations are seen 
(Croom \& Shanks 1999, hereafter CS99), as predicted by 
magnification bias due to lensing.

Four studies have looked at overdensities on scales $> 30'$, i.e.
$> 5h^{-1}$Mpc. Rodrigues-Williams and Hogan (1994) detect correlations
of Zwicky clusters and LBQS QSOs (Hewett et al. 1995) with redshifts 
$z = 1.4-2.2$ on degree angular scales. Seitz and Schneider (1995) 
extended this study to include 1 Jy QSOs. These QSOs show an association 
with Zwicky clusters with a significance of 97.7$\% $ on similar scales. 
Ferreras et al. (1997) observed a strong anti-correlation between faint 
optically selected QSOs at $z < 1.6$ near the North Galactic Pole.
However, the authors attribute this effect to selection biases associated
with identifying QSOs in crowded regions. Finally, Williams \& Irwin 
(1998, hereafter, WI98), detected overdensities of APM galaxies around 
optically selected LBQS QSOs on scales of about a degree. In particular 
they find that the galaxies primarily responsible for the detected 
overdensity are red with an average B$-$R = 2.1.

Though qualitative characteristics of these positive associations between
QSOs and galaxies are consistent with weak lensing, two explanations for 
the observations have been proposed; (1) a brightening of QSOs due to 
gravitational lensing by foreground matter overdensities, as described
above, and (2) patchy obscuration of QSOs and galaxies by intervening
Galactic dust. However, with a proper choice of background sources it is 
possible to distinguish between these two theories. For example, if dust 
obscuration is the reason for the observed galaxy overdensity around QSOs,
then a source population that is less sensitive to dust obscuration,
e.g., a radio selected QSO sample, should show less of an effect than 
one that is more sensitive to dust, like a optically selected sample.  
If lensing is the correct explanation, then such a radio sample will show 
a stronger association.

This test provides motivation for the present paper where we extend the 
work of WI98, who used optically selected QSOs, to a sample of radio 
selected QSOs. We have taken QSOs from the 1 Jy radio source catalog 
(K\"uhr et al. 1981) and searched for overdensities of APM galaxies in 
radial regions of $30'$ and $60'$ around these QSOs. In section 2, we 
discuss these catalogs along with the data selection criteria. In section 
3 we show that the correlations between 1 Jy QSOs and APM galaxies exist 
and are statistically significant. In sections 4, 5, and 6 we present 
evidence that the correlations are due to the magnification bias of weak 
lensing, and study the depence of correlations on the angular scale and 
the type of galaxies used in the analysis. Section 7 summarizes 
our findings and compares them to existing lensing models.

\section{DATA}
Our QSO sample is chosen from the 1 Jy all-sky catalog
(K\"{u}hr et al. 1981).  The catalog lists 527 radio sources 
with flux densities of $f_{5GHz} \geq 1.0$ Jy and covers 9.81 sr
of sky at Galactic latitudes $\mid b \mid  \geq \pm10^{\circ}$.  
97$\%$ of the radio sources have also been optically identified 
(Stickel et al. 1994).  We limit our sample to those QSOs with redshifts 
$\geq$ 0.5 since we would like to study the association of foreground to
background objects and not galaxies in the local environment of the 
QSOs.  For this study we have also 
chosen only those QSOs with positions on the Palomar plates
(northern hemisphere) of the APM survey. The histogram in Figure 1
shows the redshift distribution of our radio QSO sample.

The APM Catalogue (Irwin et al. 1994) is assembled from scans of the 
Palomar sky survey plates taken in the northern hemisphere and the UKST 
sky survey southern hemisphere plates. Each plate covers $\sim 6^\circ$.
The catalog lists objects detected on plates taken with 
red and blue filters  separately, along with information about each 
object's morphology and magnitude in both filters. Morphology information 
includes a classification of the object as either star-like, extended,
noise, or blended. The catalog is well calibrated internally, however, 
absolute magnitude calibrations of the data are only approximate. 
For this study the magnitude differences are not large, and our analysis 
(described below) compensates for these variations.    

The Palomar plates were taken with O and E filters.  The limiting 
magnitude for the survey is O = 21.5 (blue) and E = 20.0 (red).

We use the APM class and magnitude information to select our galaxy 
sample.  We require that the galaxies in our sample have red magnitudes 
between 18.5 and 20.0, and are classified as extended in the red. 
We make no requirement on the blue classification or magnitude of the 
objects, except that the object must be detected on the blue plates.  
We also require that our QSOs be located within $2.5^\circ$ of the 
plate center so as to reduce the effects of vignetting on the plates.  

In order to check our results, we also select a sample of Galactic stars 
with the same red and blue criteria, but that are classified as star-like 
on the red plates.
 
The average O$-$E color of galaxies in our sample is $\sim$1.8.  We 
use the color transformation of Totten \& Irwin (1998) and determine
our sample to have an average color B$-$R=1.4   This is approximately 
the color found for APM galaxies detected on the southern (UKST) plates
for galaxies with $18.5 < m_{red} < 20.0$ (WI98, Maddox et al. 1996). 
Also, the red filter of the APM northern hemisphere plates is almost
the same as the R filter of the UKST plates.  We therefore estimate our 
galaxy sample to have a redshift distribution similar to the one in WI98,
with a peak at z $=0.2$ and nearly no galaxies extending beyond z = $0.7$.

\section{ANALYSIS}
As we show below, radio selected QSOs from the 1 Jy Catalog correlate
with foreground APM galaxies. The result in not entirely surprising
as correlations have already been found between these two catalogs 
( Ben\'{\i}tez \& Mart\'{\i}nez-Gonz\'{a}lez 1995, hereafter, BM95;
Ben\'{\i}tez \& Mart\'{\i}nez-Gonz\'{a}lez 1997). The important 
difference compared to the earlier studies is that we use different
galaxy selection criteria which allows us to probe the correlations on
much larger angular scales.

Our goals in this paper are twofold. First we would like to determine
if Galactic dust obscuration is responsible for the 
observed cross correlations. Second, we would like to determine
what angular scales the signal is coming from. If the correct 
interpretation of the signal is weak lensing, the angular scale
translates into the physical size of the large scale structure 
responsible for the correlations. We also explore a related issue: 
how does the scale of correlations depend on the type of galaxies being 
used in the analysis.

As mentioned in the Introduction, the two main physical reasons for the 
cross correlation signal, namely weak lensing and dust, can be 
diffentiated by comparing radio and optically selected samples of QSOs.
Moreover, dust and lensing will produce qualitatively different trends 
when QSO-galaxy cross-correlations are studied as a function of QSO 
redshift and apparent magnitude, or limiting radio flux.
For example, dust-induced correlations are expected to be of similar
amplitude for QSOs of all redshifts, whereas lensing should 
preferentially affect QSOs in the `optimal' redshift range.

Therefore we would like to study how the correlations are affected by
QSO properties. A compact way of summarizing the 
behaviour of cross-correlations on a single angular scale is as follows. 
QSOs are divided into subsamples with limiting apparent optical magnitude,
$m_{Q,lim}$ or radio flux, $f_{Q,min}$, and a lower redshift cutoff, 
$z_{Q,min}$. Galaxies inside a circle of radius $\theta$ around each QSO 
are counted. For each QSO 100 random positions are chosen on the same APM 
plate, and at the same distance from the plate center as the QSO itself 
to minimize the effects of vignatting. The galaxy counts in these 100 
circular regions are used as comparison counts. Overdensity is then the 
ratio of the galaxy density around the real QSO and the average 
density in 100 randomly chosen circles. 

To determine the statistical significance of the correlations
in a model independent way we create 100 simulated observations, and 
compute the fraction of cases in which the over- or underdensity of
galaxies is more extreme than in the real case. For each simulated 
observation the galaxy density around a `simulated' QSO is picked from 
100 randomly chosen positions. 

To make sure that the signal is not an artifact of plate sensitivity
gradients or plate defects we repeat the entire procedure with Galactic 
stars instead of galaxies.

The results, for $30'$ and $60'$ circles are presented in Figures 2-5,
as contour plots. The top (bottom) panels are the QSO-galaxy (QSO-star) 
correlations, estimated as the overdensity of galaxies (stars) around
QSOs. The left (right) panels show overdensity (statistical significance).
Overdensity contours are at 5, 10, 20$\%$ levels for $30'$ cases, and at
5, 10$\%$ for $60'$ cases, with underdensities shown as dashed lines. 
Thickest lines represent the largest over- and underdensities.
Statistical significance is plotted at 90 and 98$\%$ confidence levels.
The shaded region in each panel marks subsamples with less than 5 QSOs.
Figures 2 and 4 use QSOs of all radio fluxes, i.e. $f_{Q,min}=1$Jy, and
show how the signal changes with apparent optical magnitude cutoff, while
Figures 3 and 5 have QSOs of all optical magnitudes, down to 
$m_V\sim 21-22$, and show how correlations behave as $f_{Q,min}$ is
changed.

All figures show a statistically significant cross correlation signal 
between QSOs and foreground galaxies, while the control correlations with 
Galactic stars do not show any significant signal. Next we discuss two
possible physical interpretations of the detected signal.

\section{PHYSICAL INTERPRETATION: DUST VS. LENSING}
All QSOs and galaxies are seen through the Galactic dust which is known 
to be patchy and extend to high Galactic latitudes (Burstein \& Heiles 
1982, Schlegel et al. 1998). Directions on the sky suffering more dust 
obscuration will show decreased counts of galaxies and QSOs thus 
leading to an apparent cross correlation between these two classes of 
objects. Even though radio selected sources will be less affected by dust 
than optically selected ones, they are still not completely free from 
dust effects because every radio selected QSO must be further 
detected and its redshift determined by optical means. 
If dust is the primary reason for the observed correlations then a radio 
selected sample should show weaker correlations than a comparably-sized
optically selected sample. 

The WI98 and the present study are well
suited for such a comparison test, as the same catalog of foreground 
galaxies, namely APM, is used in both cases. The observed situation is 
the opposite of the dust hypothesis prediction: radio selected sample 
with no optical flux
cut shows stronger correlations than the optically selected sample with
no radio flux cut, on $30'-60'$ scales, as can be seen from all the data
presented in Figures 2-5. 
These results are fully consistent with the double magnification bias 
due to lensing, which predicts that the cross correlation signal will 
become stronger if QSOs are flux-limited in more than one independent 
wavelength bands simultaneously (Borgeest et al. 1991), optical and
radio, in this case.

A further test was performed to detect lensing double magnification bias.
We repeated the calculations for Figure 2 ($m_{Q,lim}$ QSO subsamples
with no radio flux cut, $\theta=30'$) with an additional condition that 
$f_{Q,min}=1.25$Jy for all QSOs; see Figure 6.
The correlations get stronger; the $5\%$ contour is barely visible on
the right hand side of the top left panel. The average overdensity in
this plot is about $8\%$ compared to about $4\%$ in Figure 2. The
significance level has also increased, overdensities in almost all the
QSO subsamples is above $90\%$. This increase occurs despite the lower
numbers of QSOs in Figure 6 compared to Figure 2. 

Aside from magnification bias,
support for the lensing hypothesis and against dust obscuration
is presented by the behavior of the overdensity as a function of
position on the $z_{Q,min}$ vs. $m_{Q,lim}$ (or $f_{Q,min}$) plane.
The strongest correlations are seen for QSOs at intermediate redshifts 
($z_{Q,min} \sim 1.0$). This is the optimal location of sources for 
lenses at $z_l\sim 0.1-0.3$.
On the other hand, Galactic dust should not be able to differentiate 
between QSOs based on their redshifts. Also, the effects of single
wavelength band magnification bias are apparent from Figure 2-5: in any
given figure correlations get stronger for brighter QSOs. 

We conclude that the cross-correlation signal between 1 Jy QSOs and APM
galaxies in our sample is due to weak lensing of QSO by the galaxies and 
associated dark matter. 

Additional independent confirmation of the lensing origin of the 
QSO-galaxy associations in general comes from a recent study by 
CS99 who detect anticorrelations 
between faint optically selected QSOs and foreground groups of galaxies. 
Because these authors used QSO and galaxy samples that are different 
from ours we cannot directly compare our respective results. However
if all the existing QSO-galaxy association observations are to be
explained by a single process then the combination of the results
presented here, in WI98 and in CS99 rule out the dust hypothesis. 
For positive correlation
results, as seen in this work, one invokes dust foreground to both
QSOs and galaxies, i.e.  Galactic dust, 
to make both QSOs and galaxies overdense in some regions of the sky
and underdense in others. For anticorrelation results, as detected in
CS99, one needs to invoke
dust {\it intrinsic} to the galaxy groups and clusters 
to obscure QSOs only in the directions of the lenses. Thus two
very different types of dust are needed to explain the two phenomena,
while lensing magnification bias accounts for both types of observations:
anticorrelations are predicted with QSO samples flux-limited below the 
turnover in the number counts and positive correlations are predicted
for samples limited above the turnover.

\section{ANGULAR SCALE OF CORRELATIONS }
It is apparent from Figures 2-5 that the amplitude of the cross 
correlation signal 
decreases with the angular scale. For example, the representative
overdensity in Figure 3 ($\theta=30'$) is $10\%$, and drops to $6-7\%$ on
Figure 5 ($\theta=60'$). Note that the statistical significance stays 
roughly the same on both scales. If the signal originates entirely from 
small scales, $<30'$, then one would expect the galaxy overdensity to 
scale as $\sim 1/\theta^2$. 
The observations indicate a much slower decline, 
$\sim 1/\sqrt{\theta}$, implying that the signal is not limited to 
$<30'$, but an appreciable contribution arises from $\sim$degree scales.

Ideally, we would like to extend the type of analysis carried out in the
previous Sections to scales smaller, and larger than $30'-60'$. 
However, on scales larger than about a degree we run into a problem with
APM plate boundaries because the useful area on each plate is 
limited to $2.5^\circ$ around the plate center. To look for correlations 
on scales substantially larger than a degree one would need to match 
adjacent plates, a task too uncertain when the signal 
being sought after is of the order of a few percent at best.

On scales smaller than about $30'$ we run into a different problem. The
numbers of galaxies drop significantly as the circular area around each
QSO is decreased. Because there are not that many QSOs to start with
splitting them futher into $m_{Q,lim}$ or $f_{Q,min}$, and $z_{Q,min}$
subsamples is not feasible. 
To look for signal on smaller scales we combine QSOs into much larger
subsamples and plot radial density gradients. Using our full QSO sample,
we find no radial galaxy gradients around QSOs on these small scales. 
Thus we conclude that the galaxies selected as
having $18.5<m_{red}<20$, with no color cut are tracing the large scale
structure on $\sim 10h^{-1}$ Mpc scales, but not the more compact 
structures on $\sim 1-2h^{-1}$Mpc scales.

\section{GALAXY SELECTION CRITERIA AND CORRELATIONS}
The lack of radial gradient of galaxies within $30'$ around 1 Jy QSOs 
is in apparent contradiction with the results of BM95 who detect strong 
correlations on $\theta<10'$, but see no signal beyond that scale.  

We suggest that the discrepancy arises because  BM95 and the 
present work use different galaxy selection criteria and thereby are 
studing lensing properties of different populations of galaxies, which 
apparently trace dark matter on different scales.

The two populations are primarily distinguished by their angular size
and apparent magnitudes. Our galaxies have $18.5<m_{red}<20$ and can have
such small angular sizes that they appear unresolved on one set of APM 
plates (blue) As a result a non-negligible fraction of galaxies in this 
study, and those in WI98 do appear point-like on the blue plates. (WI98 
find that these objects contribute significantly to the detected signal.)
BM95 select galaxies with $m_{red}<19.5$, and insist that they have a large
enough angular extent to be resolved on both blue and red plates. 
Furthermore, the average angular size of BM95 galaxies, as indicated by
the semi-major axes of their images on the APM plates is about
1.5 times larger than those of our galaxies. Thus, BM95 use galaxies 
that are roughly 1.5 times bigger and about 4 times brighter than
those in our sample. 

Assuming that the two sets of galaxy populations are at the same redshift,
$z_l\sim 0.1-0.3$, and remembering that the correlations in the present
study extend to a $\sim$ degree, while those in BM95 are confined to
$< 10'$, we are led to conclude that intrinsically brighter, bigger galaxies
trace compact, $\sim 1-2h^{-1}$ Mpc mass concentrations, while fainter
smaller galaxies trace more extended, smaller contrast mass concentrations
on $\sim 10h^{-1}$ Mpc scales. This is, of course, just a speculation;
more evidence is needed to support this claim. Other possibilities 
should also be considered. For example, the two populations of galaxies
can be at different redshifts, with BM95 galaxies being twice as close
to the observer as those in the present study. If BM95 galaxies are 
relatively nearby, say at a typical redshift of $\simlt 0.1$, then 
difference in $\Sigma_{crit}$, critical surface mass density for lensing
between the location of BM95 galaxies and those in the present work will
have to be taken into account.



It is worth noting that regardless of the angular scale of associations
the galaxies primarily responsible for the signal are red, with B$-$R
greater than about 2 (WI98; BM95; 
Ben\'{\i}tez \& Mart\'{\i}nez-Gonz\'{a}lez 1997). These red 
galaxies are probably mostly ellipticals, whose distribution is believed 
to be more biased with respect to mass than that of bluer, presumably 
mostly spiral galaxies.

\section{CONCLUSIONS AND DISCUSSION}

In this paper we have extended the work of WI98 to a sample of 
1 Jy radio selected QSOs.  We have searched for and found correlations
of these radio QSOs and red APM galaxies on scales of $30\arcmin$ 
and $60\arcmin$. We demonstrate that observed correlations are 
qualitatively consistant with a gravitational lensing explanation and
are inconsistant with a dust obscuration model. The detected overdensity  
of galaxies around these radio selected QSOs is larger than that found 
in WI98 for a sample of optically selected QSOs, on similar angular 
scales. We ascribe the difference to double magnification bias.

For our galaxy sample we find no radial gradients in galaxy density
around QSOs on $< 30'$ scales. This is in contradiction with the work of 
BM95, who also use 1 Jy QSOs and APM galaxies. 
We suggest that the discrepancy is 
due to differences in galaxy sample selection. The difference in selection 
criteria translate, on the average, into BM95 galaxies being about 1.5 
times bigger in radius and about 4 times brighter than those in our
current sample. We speculate that B95 population of galaxies are
better tracers of compact $\sim 1-2h^{-1}$ Mpc structures at 
$z\sim 0.1- 0.3$, while our galaxies trace more extended structures,
$\sim 10h^{-1}$ Mpc in a similar redshift. 

In this paper we have argued that the  most likely explanation of the
correlations is weak gravitational lensing. We have shown that 
qualitatively the correlations follow lensing expectations. What does 
not follow the predictions of standard lensing is the amplitude of these 
correlations.

The amplitude of correlations found in this work is greater than that 
in WI98, which already exceed the predictions of theoretical models.
As dicussed extensively in WI98, the analytical models of Dolag and 
Bartelmann (1997) and Sanz et al. (1997) and phenomenological predictions
using the observed properties of APM galaxies under-predict the 
amplitude of the overdensity on degree scales by a factor of $5-10$.
In order to bring the models in line with the observations, assuming that
our model of light propagation through the universe is correct, requires 
a change in the important parameters that describe either the sources or
the foreground lenses.  WI98 consider two possiblities in some detail. 
Either, (1) the source population has very steep magnitude-number counts
at $z > 1$, or (2) galaxies are biased low with respect to mass, i.e. 
$\sigma_8$, the
rms dispersion in mass within $R=8h^{-1}$Mpc spheres, is much larger than 
one.  WI98 point out that neither of these options is likely in light of 
other observations and that, likewise, appropriate combinations of these 
two explanations lead to numbers which are also out of the acceptable 
range of the parameter space. Currently we have no explanation for the 
disagreement of the models and observations. Clearly, more work is
required both on observational and theoretical fronts to tackle this
problem.

\acknowledgments{The authors would like to thank M.J. Irwin for his extensive help in using the APM catalog and for all his very useful comments. D.J.N.  acknowledges support from NASA GSRP S96-UMF-003 and the Danforth-Compton Fellowship. }

\begin{figure}
       \figurenum{1}
       \epsscale{1}
       \plotone{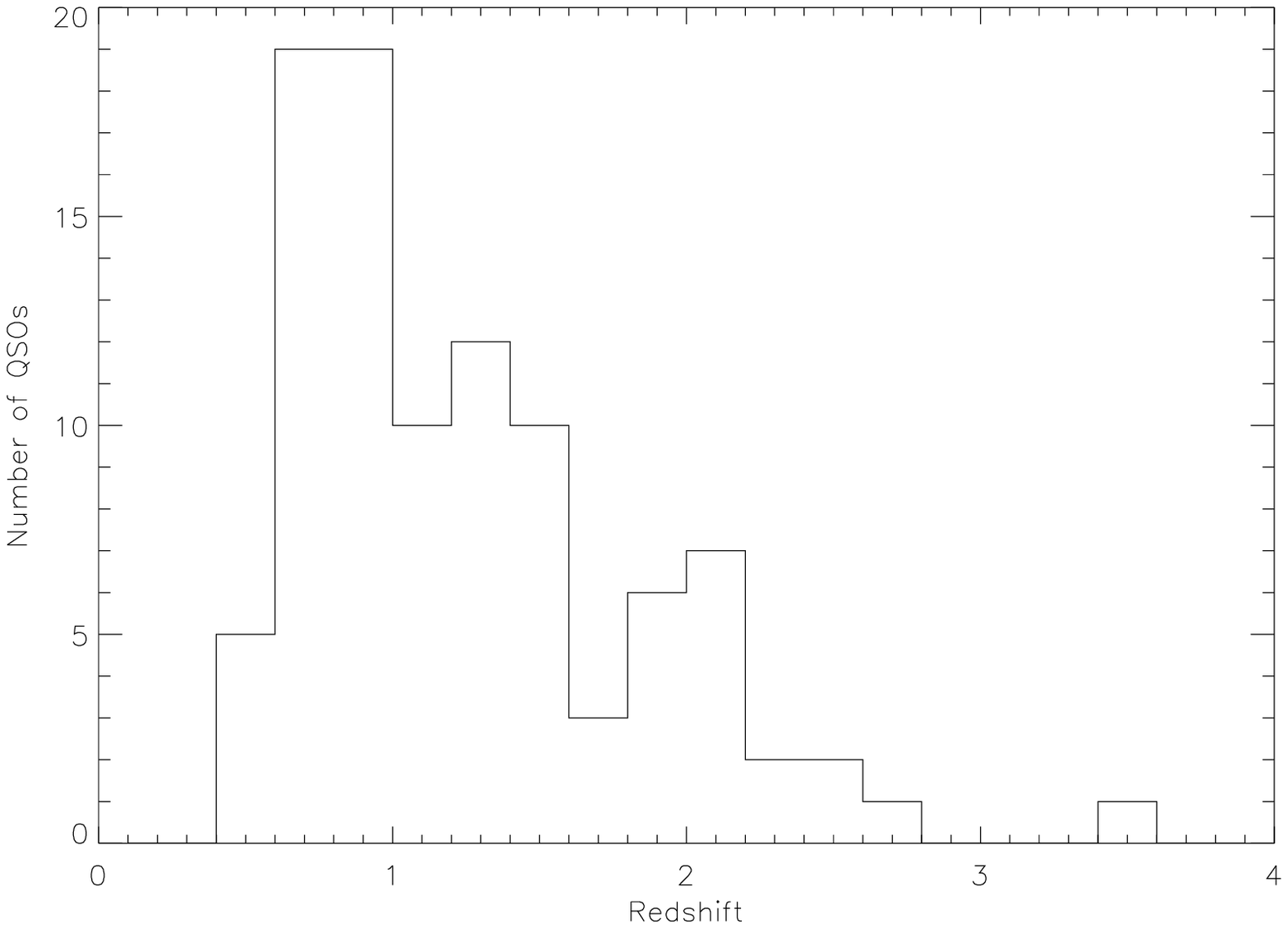}
        \caption{Histogram of the number of 1 Jy QSOs versus redshift 
used in this study.   \label{f1}}
\end{figure} 

\begin{figure}
       \figurenum{2}
       \epsscale{1}
       \plotone{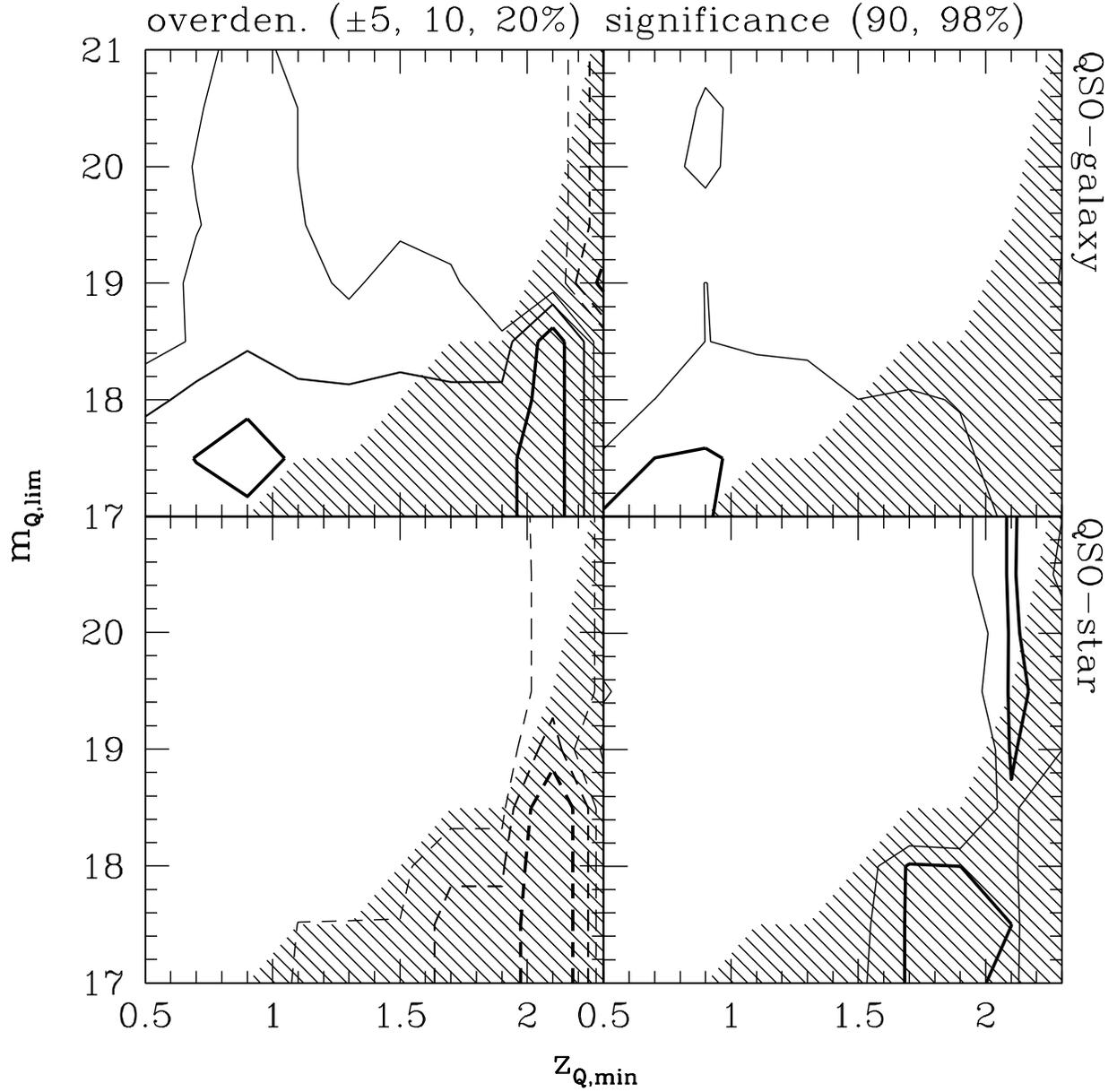}
        \caption{Overdensity and statistical significance contour plots for
$30\arcmin$ radial regions around QSOs with $f_{Q,min} = 1.0 $Jy  as the limiting optical magnitude changes. The top (bottom) panels are the QSO-galaxy (QSO-star) correlations, estimated as the overdensity of galaxies (stars) around QSOs. The left (right) panels show overdensity (statistical significance). Shaded regions have less than 5 QSOs contributing to the estimate. (See text for further details.)   \label{f2}}
\end{figure}

\begin{figure}
       \figurenum{3}
       \epsscale{1}
       \plotone{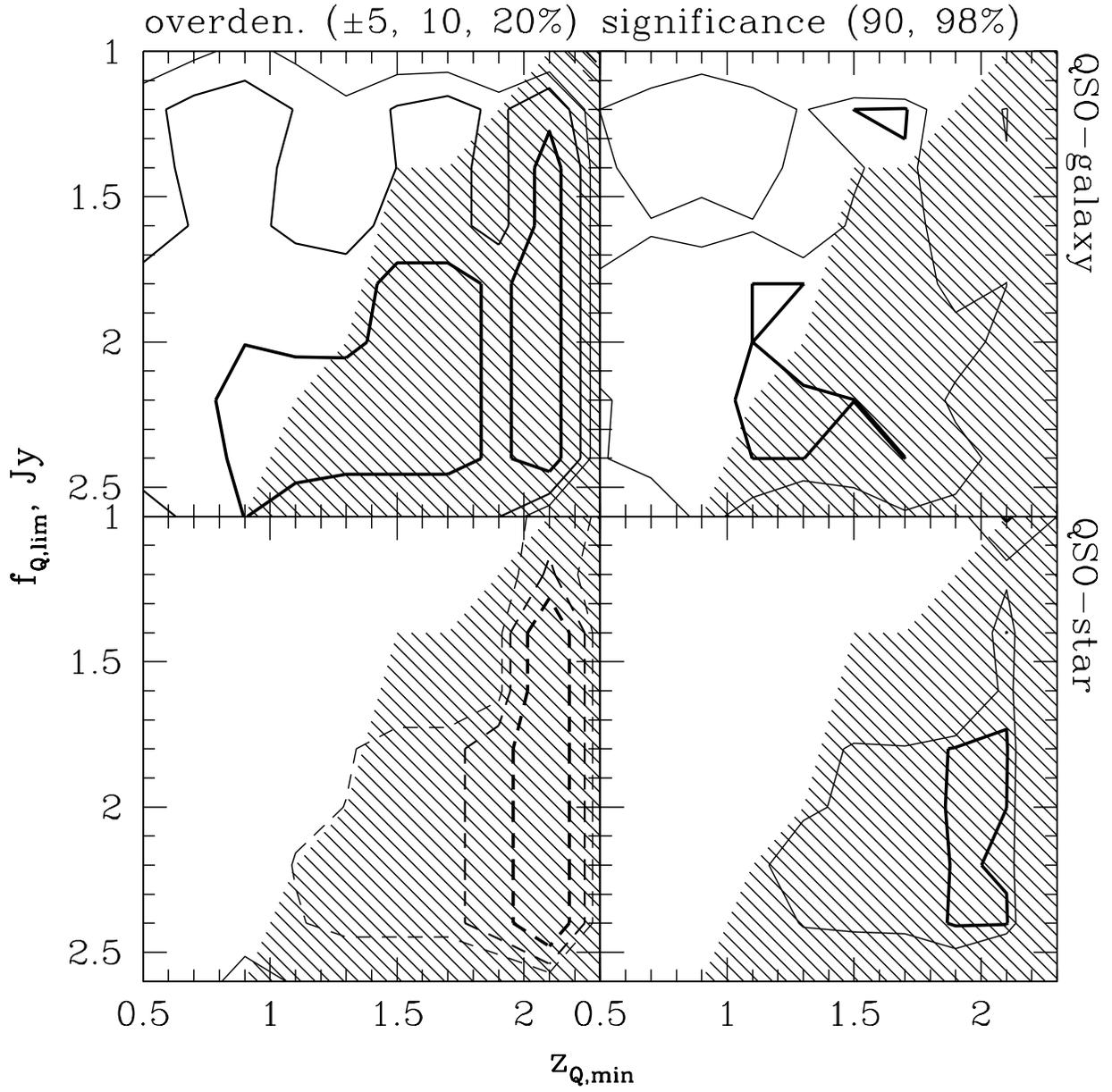}
        \caption{Overdensity and statistical significance contour plots similar
to Figure 2 except that all optical magnitudes are used and $f_{Q,min}$ 
changes.   \label{f3}}
\end{figure}

\begin{figure}
       \figurenum{4}
       \epsscale{1}
       \plotone{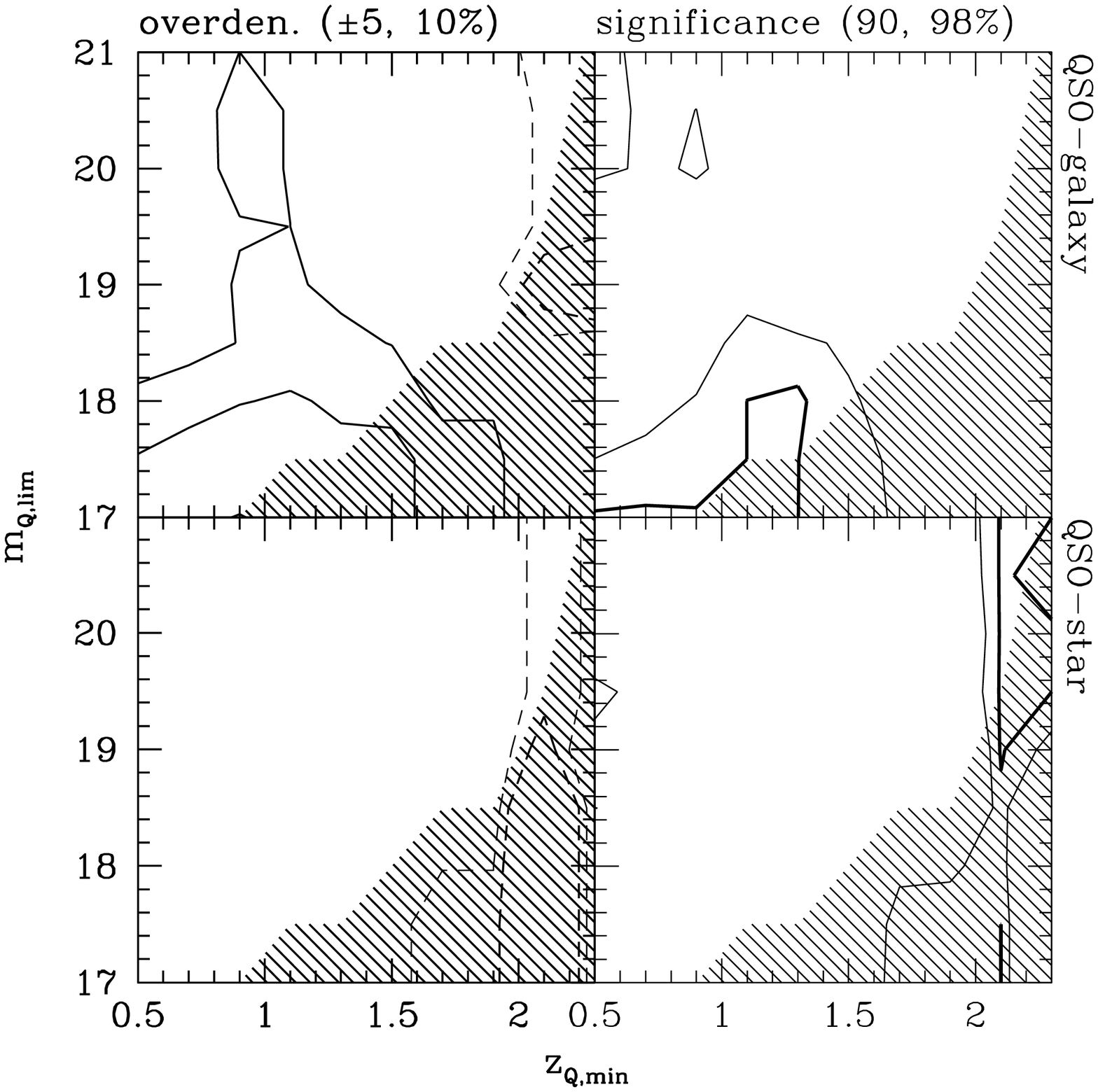}
        \caption{Same as Figure 2, but for $60\arcmin$ radial regions.    \label{f4}}
\end{figure}

\begin{figure}
       \figurenum{5}
       \epsscale{1}
       \plotone{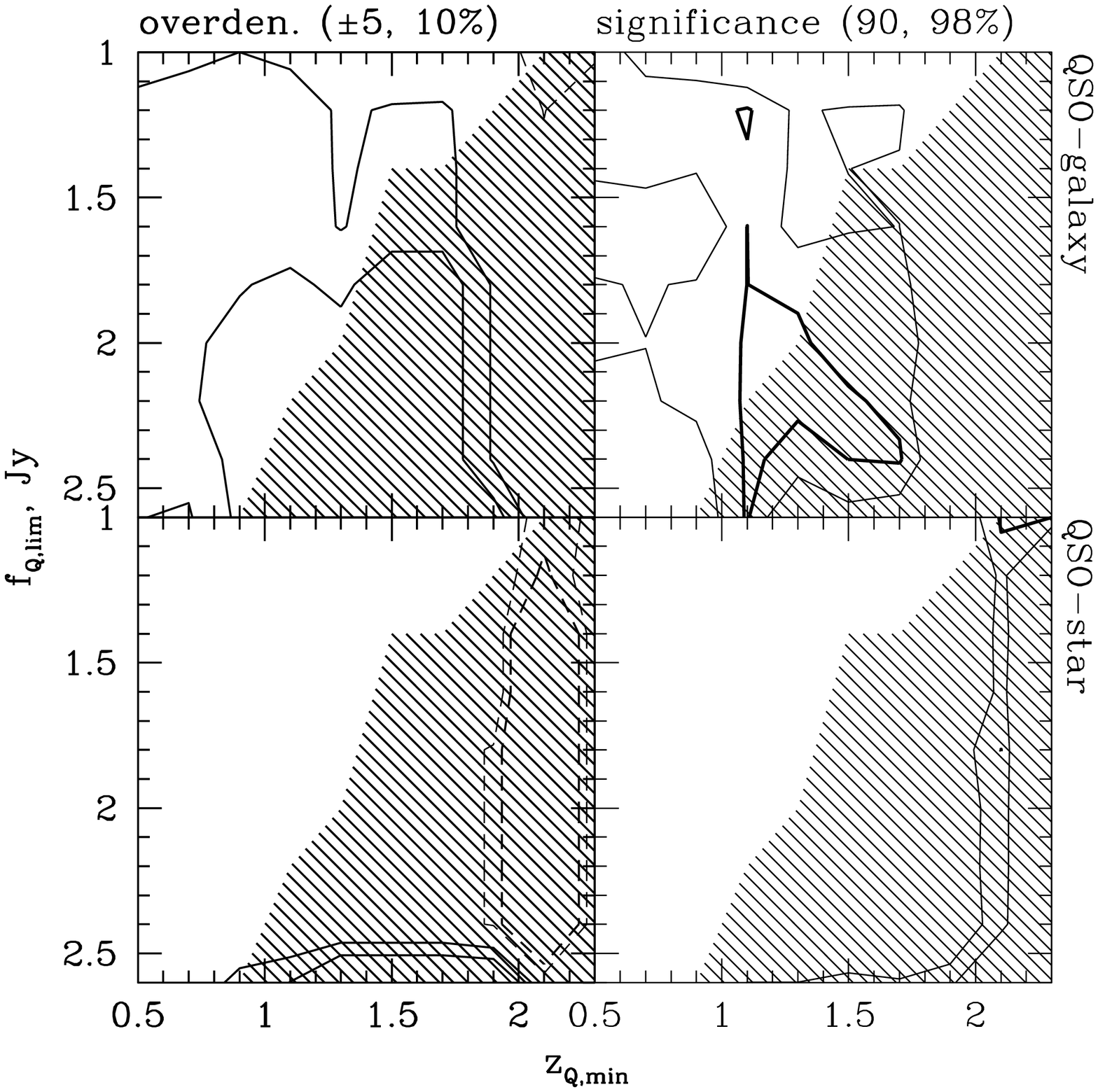}
        \caption{Same as Figure 3, but for $60\arcmin$ radial regions. \label{f5}}
\end{figure}

\begin{figure}
       \figurenum{6}
       \epsscale{1}
       \plotone{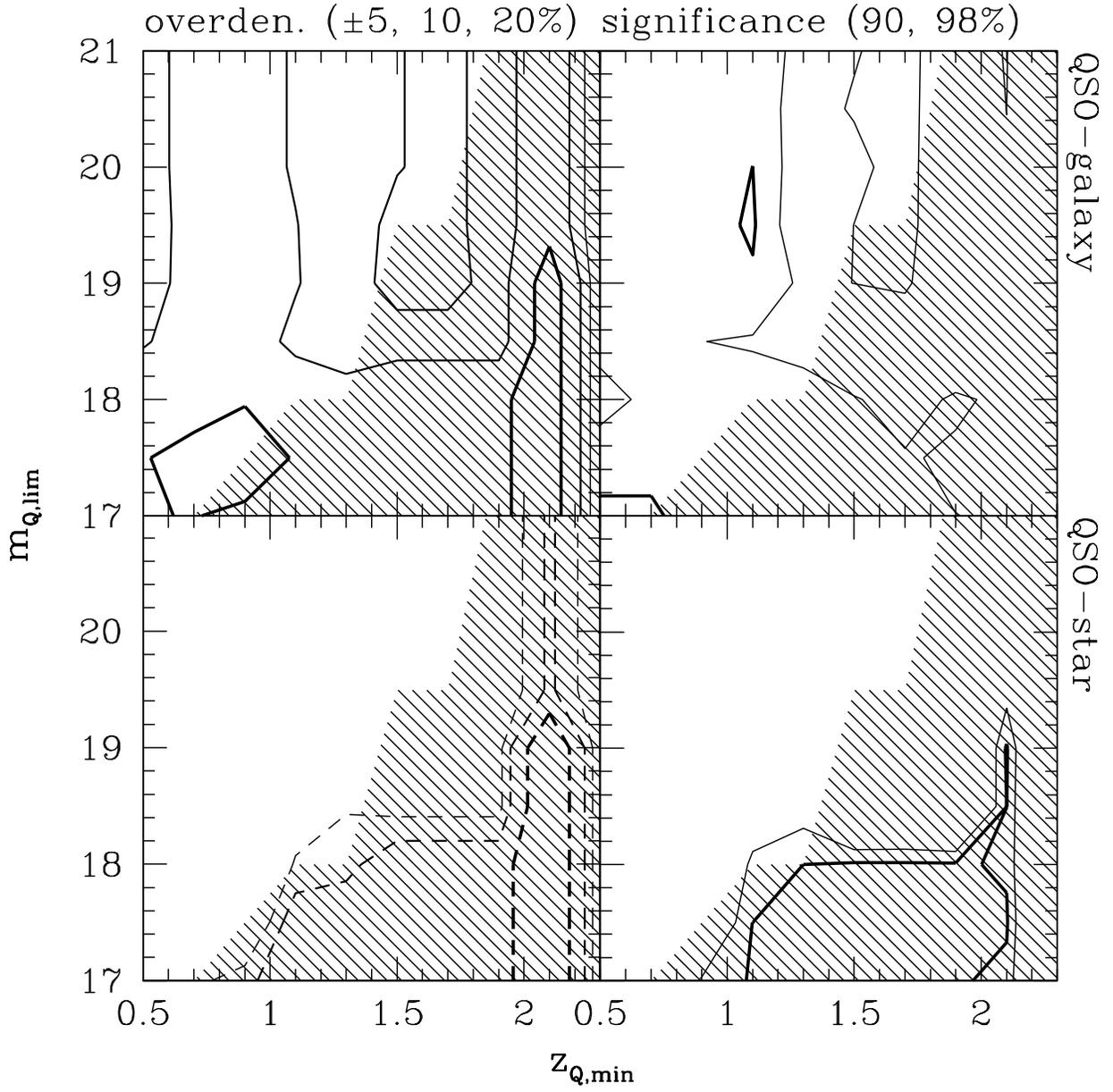}
        \caption{Overdensity and statistical significance contour plots similar
to Figure 2 except that $f_{Q,min} = 1.25$ Jy  as the limiting optical
 magnitude changes.   \label{f6}}
\end{figure}

\end{document}